# Quasiparticle relaxation rate and shear viscosity of superfluid $^3He$-$A_1$ at low temperatures

M. A. Shahzamanian and R. Afzali

Department of Physics, Faculty of Sciences, University of Isfahan, 81744 Isfahan, Iran
E-mail addresses: shahzamanian@hotmail.com
afzali2001@hotmail.com

Abstract

Quasiparticle relaxation rate, $\tau_p^{-1}$, and the shear viscosity tensor of the $A_1$−phase of superfluid $^3He$ are calculated at low temperatures and melting pressure, by using Boltzmann equation approach in momentum space. The collision integral is written in terms of inscattering and outscattering collision integrals. The interaction between normal and Bogoliubov quasiparticles is considered in calculating transition probabilities in the binary, decay and coalescence processes. We obtain that both $\tau_{p\uparrow}^{-1}$ and $\tau_{p\downarrow}^{-1}$ are proportional to $T^2$. The shear viscosities $\eta_{xy}$, $\eta_{xz}$ and $\eta_{zz}$ are proportional to $(T/T_c)^{-2}$. The constant of proportionality of the shear viscosity tensor is in nearly good agreement with the experimental results of Roobol et al., and our exact theoretical calculation.

In this paper, we deal with superfluid $^3He - A_1$ at low temperatures, melting pressure and high magnetic field up to 15 $T$. We assume that magnetic field is sufficiently high so that all quasiparticles with spin up go to superfluid state and all quasiparticles with spin down stay in normal state. Also we consider normal-superfluid interactions that come from the scattering between superfluid quasiparticles in the spin-up population, the so-called Bogoliubov quasiparticles and the normal fluid quasiparticles in the spin-down population. In a normal Fermi liquid at low temperatures the only important collision process is the binary scattering of quasiparticles, but in a superfluid the quasiparticle number is not conserved and other processes as well as binary processes can occur. So we take into account these processes such as decay processes in which one Bogoliubov quasiparticle decays into three and coalescence processes in which three Bogoliubov quasiparticles coalesce to produce one.

The interaction term in the Hamiltonian of system is

$$H = \frac{1}{4}\sum_{1,2,3,4}\langle 3,4|T|1,2\rangle a^\dagger_4 a^\dagger_3 a_1 a_2 \qquad (1)$$

where i = 1,2,3 and 4, stands for both momentum ($\vec{p}_i$) and spin ($\sigma_i$) variables. By using Bogoliubov transformation, the quasiparticle creation $a^\dagger_{\vec{p},\uparrow}$ and annihilation $a_{\vec{p},\uparrow}$ operators may be replaced by the creation and annihilation operators $\alpha^\dagger_{\vec{P},\sigma}$ and $\alpha_{\vec{P},\sigma}$ in the superfluid. The creation and annihilation operators with spin down in $H$ are left intact. Bogoliubov transformation may be written as

$$a_{\vec{p},\sigma} = u_{\vec{p},\sigma\sigma'}\alpha_{\vec{p},\sigma'} - v_{\vec{p},\sigma\sigma'}\alpha^{\dagger}_{-\vec{p},\sigma'} \quad , \quad a^{\dagger}_{\vec{p},\sigma} = v^{*}_{\vec{p},\sigma\sigma'}\alpha_{\vec{p},\sigma'} + u_{\vec{p},\sigma\sigma'}\alpha^{\dagger}_{-\vec{p},\sigma'} \tag{2}$$

where

$$u_{\vec{p},\sigma\sigma'} = \left[\frac{1}{2}\left(1+\frac{\varepsilon_{\vec{p}}}{E_{\vec{p}}}\right)\right]^{1/2}\delta_{\sigma\sigma'} \quad , \quad v_{\vec{p},\sigma\sigma'} = \left[\frac{1}{2}\left(1-\frac{\varepsilon_{\vec{p}}}{E_{\vec{p}}}\right)\right]^{1/2}\delta_{\sigma\sigma'} \quad , \quad E_{\vec{p}} = \left(\varepsilon_{\vec{p}}^{2} + \left|\Delta_{\vec{p}\uparrow\uparrow}\right|^{2}\right)^{1/2}\text{sgn}\,\varepsilon_{\vec{p}}$$

with $v_{-\vec{p},\uparrow\uparrow} = -v_{\vec{p},\uparrow\uparrow}$ and $u_{-\vec{p},\uparrow\uparrow} = u_{\vec{p},\uparrow\uparrow}$. $\varepsilon_{\vec{p}}$ and $\Delta_{\vec{p}\uparrow\uparrow}$ are the normal-state quasiparticle energy measured with respect to the chemical potential and the magnitude of the gap in the direction $\vec{p}$ on the Fermi surface respectively [1]. $\left|\Delta_{\vec{p}\uparrow\uparrow}\right|$ is equal to $\Delta(T)\sin\theta_{p}$, where $\Delta(T)$ is the maximum gap and $\theta_{p}$ is the angle between the quasiparticle momentum and gap axis $\hat{\ell}$ that suppose to be in the direction of z-axis. Then by using Eq.(2) in Eq.(1) we have

$$H = \frac{1}{4}\sum_{\vec{p}_1,\vec{p}_2,\vec{p}_3,\vec{p}_4}\left\{\left[<3\uparrow 4\uparrow|T|1\uparrow 2\uparrow>\left(u_{4\uparrow\uparrow}\alpha^{\dagger}_{4\uparrow}-v^{*}_{4\uparrow\uparrow}\alpha_{-4\uparrow}\right)\left(u_{3\uparrow\uparrow}\alpha^{\dagger}_{3\uparrow}-v^{*}_{3\uparrow\uparrow}\alpha_{-3\uparrow}\right)\left(u_{1\uparrow\uparrow}\alpha_{1\uparrow}-v_{1\uparrow\uparrow}\alpha^{\dagger}_{-1\uparrow}\right)\left(u_{2\uparrow\uparrow}\alpha_{2\uparrow}-v_{2\uparrow\uparrow}\alpha^{\dagger}_{-2\uparrow}\right)\right]$$

$$+\left[\langle 3\downarrow 4\downarrow|T|1\downarrow 2\downarrow\rangle a^{\dagger}_{4\downarrow}a^{\dagger}_{3\downarrow}a_{1\downarrow}a_{2\downarrow}\right]$$

$$+\left[\langle 3\downarrow 4\uparrow|T|1\uparrow 2\downarrow\rangle\left(u_{4\uparrow\uparrow}\alpha^{\dagger}_{4\uparrow}-v^{*}_{4\uparrow\uparrow}\alpha_{-4\uparrow}\right)a^{\dagger}_{3\downarrow}\left(u_{1\uparrow\uparrow}\alpha_{1\uparrow}-v_{1\uparrow\uparrow}\alpha^{\dagger}_{-1\uparrow}\right)a_{2\downarrow}\right]$$

$$+\left[\langle 3\uparrow 4\downarrow|T|1\uparrow 2\downarrow\rangle a^{\dagger}_{4\downarrow}\left(u_{3\uparrow\uparrow}\alpha^{\dagger}_{3\uparrow}-v^{*}_{3\uparrow\uparrow}\alpha_{-3\uparrow}\right)\left(u_{1\uparrow\uparrow}\alpha_{1\uparrow}-v_{1\uparrow\uparrow}\alpha^{\dagger}_{-1\uparrow}\right)a_{2\downarrow}\right]$$

$$+\left[\langle 3\downarrow 4\uparrow|T|1\downarrow 2\uparrow\rangle\left(u_{4\uparrow\uparrow}\alpha^{\dagger}_{4\uparrow}-v^{*}_{4\uparrow\uparrow}\alpha_{-4\uparrow}\right)a^{\dagger}_{3\downarrow}a_{1\downarrow}\left(u_{2\uparrow\uparrow}\alpha_{2\uparrow}-v_{2\uparrow\uparrow}\alpha^{\dagger}_{-2\uparrow}\right)\right]$$

$$+\left[\langle 3\uparrow 4\downarrow|T|1\downarrow 2\uparrow\rangle a^{\dagger}_{4\downarrow}\left(u_{3\uparrow\uparrow}\alpha^{\dagger}_{3\uparrow}-v^{*}_{3\uparrow\uparrow}\alpha_{-3\uparrow}\right)a_{1\downarrow}\left(u_{2\uparrow\uparrow}\alpha_{2\uparrow}-v_{2\uparrow\uparrow}\alpha^{\dagger}_{-2\uparrow}\right)\right]\right\} \tag{3}$$

The transition probabilities for decay process, for example, may be written as $W_{13}(\downarrow\uparrow) = 2\pi\left|\langle\vec{p}_3\uparrow,\vec{p}_4\uparrow,-\vec{p}_2\downarrow|H|\vec{p}_1\downarrow\rangle\right|^2$ (throughout of this paper we put $\hbar \equiv K_B \equiv 1$) and similarly for other processes. By using Wick's theorem after straightforward lengthy calculations we get

$$W_{22}(\downarrow\downarrow) = 2\pi\left|T_{t_I}\right|^2$$

$$W_{22}(\uparrow\uparrow) =$$

$$2\pi\left\{\left[\left(|v_1|^2|v_2|^2|v_3|^2|v_4|^2 + |u_1|^2|u_2|^2|u_3|^2|u_4|^2 + 2u_1v_1u_2v_2u_3v_3u_4v_4\right)\left|T_{t_I}\right|^2\right]\right.$$

$$+\left[\left(|u_1|^2|v_2|^2|u_3|^2|v_4|^2 + |v_1|^2|u_2|^2|v_3|^2|u_4|^2 + 2u_1v_1u_2v_2u_3v_3u_4v_4\right)\left|T_{t_{II}}\right|^2\right]$$

$$+\left[\left(|v_1|^2|u_3|^2|u_2|^2|v_4|^2 + |v_2|^2|u_4|^2|u_1|^2|v_3|^2 + 2u_1v_1u_2v_2u_3v_3u_4v_4\right)\left|T_{t_{III}}\right|^2\right]$$

$$-\left[\left(|v_1|^2|v_4|^2 u_2v_2u_3v_3 + |v_2|^2|v_3|^2 u_1v_1u_4v_4 + |u_2|^2|u_3|^2 u_1v_1u_4v_4 + |u_1|^2|u_4|^2 u_2v_2u_3v_3\right)\left(T_{t_I}^{*}T_{t_{III}} + T_{t_I}T_{t_{III}}^{*}\right)\right]$$

$$-\left[\left(|v_4|^2|v_2|^2 u_1v_1u_3v_3 + |v_1|^2|v_3|^2 u_2v_2u_4v_4 + |u_1|^2|u_3|^2 u_2v_2u_4v_4 + |u_2|^2|u_4|^2 u_1v_1u_3v_3\right)\left(T_{t_I}^{*}T_{t_{II}} + T_{t_I}T_{t_{II}}^{*}\right)\right]$$

$$+\left[\left(|u_3|^2|v_4|^2 u_1v_1u_2v_2 + |v_1|^2|u_2|^2 u_3v_3u_4v_4 + |u_1|^2|v_2|^2 u_3v_3u_4v_4 + |v_3|^2|u_4|^2 u_1v_1u_2v_2\right)\left(T_{t_{III}}^{*}T_{t_{II}} + T_{t_{III}}T_{t_{II}}^{*}\right)\right]\right\}$$

$$W_{13}(\uparrow\uparrow) = 2\pi\left|\left[\left(v_1^{*}u_2v_3v_4 - u_1v_2u_3u_4\right)T_{t_I} + \left(v_1^{*}v_2v_3u_4 - u_1u_2u_3v_4\right)T_{t_{II}} + \left(v_1^{*}v_2u_3v_4 - u_1u_2v_3u_4\right)T_{t_{III}}\right]\right|^2$$

$$W_{31}(\uparrow\uparrow) = 2\pi\left|\left[\left(v_1^{*}v_2^{*}u_3v_4 - u_1u_2v_3^{*}u_4\right)T_{t_I} + \left(u_1v_2^{*}v_3^{*}v_4 - v_1^{*}u_2u_3v_4\right)T_{t_{II}} + \left(v_1^{*}u_2v_3^{*}v_4 - u_1v_2^{*}u_3u_4\right)T_{t_{III}}\right]\right|^2$$

$W_{22}(\uparrow\downarrow) =$

$(\pi/2)\{|u_1|^2 |u_3|^2 |(-T_{s_I} + T_{t_I})|^2 + |v_1|^2 |v_3|^2 |(-T_{s_{II}} + T_{t_{II}})|^2 + |u_1|^2 |u_4|^2 |(T_{s_I} + T_{t_I})|^2 + |v_1|^2 |v_4|^2 |(T_{s_{III}} + T_{t_{III}})|^2$

$-u_1^* v_1^* u_3 v_3 (-T_{s_I} + T_{t_I})^* (-T_{s_{II}} + T_{t_{II}}) - u_1 v_1 u^*_3 v^*_3 (-T_{s_I} + T_{t_I})(-T_{s_{II}} + T_{t_{II}})^*$

$-u_1^* v_1^* u_4 v_4 (T_{s_I} + T_{t_I})^* (T_{s_{III}} + T_{t_{III}}) - u_1 v_1 u^*_4 v^*_4 (T_{s_I} + T_{t_I})(T_{s_{III}} + T_{t_{III}})^*\}$

$W_{13}(\downarrow\uparrow) = (\pi/2)\{ |u_3|^2 |v_4|^2 |T_{s_{II}} + T_{t_{II}}|^2 + |v_3|^2 |u_4|^2 |-T_{s_{III}} + T_{t_{III}}|^2$

$+ u_3 v_3 u_4^* v_4^* (T_{s_{II}} + T_{t_{II}})^* (-T_{s_{III}} + T_{t_{III}}) + u_3^* v_3^* u_4 v_4 (T_{s_{II}} + T_{t_{II}})(-T_{s_{III}} + T_{t_{III}})^* \}$

$W_{31}(\downarrow\uparrow) = (\pi/2)\{ |v_3|^2 |u_4|^2 |-T_{s_{III}} + T_{t_{III}}|^2 + |u_3|^2 |v_4|^2 |T_{s_{II}} + T_{t_{II}}|^2$

$+ u_3 v_3 u_4^* v_4^* (-T_{s_{III}} + T_{t_{III}})^* (T_{s_{II}} + T_{t_{II}}) + u_3^* v_3^* u_4 v_4 (-T_{s_{III}} + T_{t_{III}})(T_{s_{II}} + T_{t_{II}})^* \}$

(4)

Also $W_{22}(\downarrow\uparrow)$ is obtained by replacing $1 \leftrightarrow 2$ and $3 \leftrightarrow 4$ in $W_{22}(\uparrow\downarrow)$. $T_s$ and $T_t$ are the scattering amplitudes for pairs of quasiparticles in singlet and triplet states respectively. By using Pfitzner procedure [2], $T_{t_I}$ and $T_{s_I}$ are given by

$$N(0)T_{t_I,s_I}(\nu,P) = \sum_{k=0}^{\infty} \sum_{l=0}^{k} a_{lk}(k+1)^{1/2}(2l+1)^{1/2}(P^2/4-1)^l P_l(\nu) P_{k-l}^{(2l+1,0)}(P^2/2-1)$$

$k = 0,1,...,$ ; $l = 0,1,...,k$

(5)

The coefficients with $l$ even (odd) belong to the singlet (triplet) part of the quasiparticle scattering amplitude (QSA). $N(0) = m^* p_F / \pi^2$, $P_l(\nu)$ and $P_n^{(a,b)}(x)$ are the density of states at the Fermi level, the Legendre polynomials and the Jacobi polynomials respectively. Definitions of $P$ and $\nu$ are $P \equiv 2\cos(\theta/2)$ and $\nu \equiv \cos\varphi$ where $\theta$ is the angle between the momenta of the incoming particles namely $\vec{p}_1$ and $\vec{p}_2$, and $\varphi$ is the angle between the planes spanned by the momentum vectors of the incoming particles and the outgoing particles. Similar equations for $T_{t_{II}}$ and $T_{s_{II}}$ ( $T_{t_{III}}$ and $T_{s_{III}}$ ) with replacement $\theta$ by $\theta_{II}$ ( $\theta_{III}$ ) and $\varphi$ by $\varphi_{II}$ ( $\varphi_{III}$ ) can be used. $\theta_{II}$ and $\varphi_{II}$ ($\theta_{III}$ and $\varphi_{III}$) are related to $\theta$ and $\varphi$ by the following equations when $a = 1$ ($a = -1$):

$$cos\theta_{II,III} = -\cos^2\frac{\theta}{2} + a(1-\cos^2\frac{\theta}{2})\cos\varphi \quad , \quad cos\varphi_{II,III} = \frac{(1-\cos^2\frac{\theta}{2})\cos\varphi + a\left(3\cos^2\frac{\theta}{2}-1\right)}{a(-1+\cos^2\frac{\theta}{2})\cos\varphi + 1 + \cos^2\frac{\theta}{2}}$$

(6)

At low temperatures, we have $\sin\theta_{p_i} \simeq 0$ $(i=1,2,3,4)$ [3], then we may write $\nu \simeq 0$ and $u \simeq 1$. It is noted that if we apply other approximations for $\nu$ and $u$, the temperature dependence does not change, however some small changes appear in the numerical coefficients of the transition probabilities. By substituting the values of $\nu$ and $u$ in the equations of the transition probabilities' (Eq. (4)), we obtain

$W_{22}(\downarrow\downarrow) = 2\pi |T_{t_I}|^2 \quad , \quad W_{22}(\uparrow\uparrow) \simeq 2\pi |T_{t_I}|^2 + (\pi/4)(|T_{t_{II}}|^2 + |T_{t_{III}}|^2)$

$W_{22}(\uparrow\downarrow) = W_{22}(\downarrow\uparrow) \simeq (\pi/2)(|-T_{s_I} + T_{t_I}|^2 + |T_{s_I} + T_{t_I}|^2)$

(7)

and the other transition probabilities have nearly a zero value. It is noted that only binary processes are dominated at low temperatures, and this is also the case for calculating the thermal diffusion coefficient of the A-phase at low temperatures [3]. We note that $\theta$ is small for the superfluid case and its maximum value is $\pi T/\Delta(0)$ [3] where maximum gap, $\Delta(0)$, due to strong coupling effects is equal to $1.77T_c$. To make clear the reason for the smallness of $\theta$, we note the following points. First, at low temperatures, Bogoliubov quasiparticle momentum vectors are located around the nodes of energy gap; consequently, these vectors make small angles around the gap axis. Second, for Bogoliubov quasiparticles, $\theta$ follows $\theta_p$ in terms of being small. Therefore by using Eqs. (5)-(6), $T_{s_I}$, $T_{t_I}$, $T_{t_{II}}$ and $T_{t_{III}}$ in Eq. (7) explicitly are

$$N(0)T_{s_I} = 2.47 + 6.61\cos^2\frac{\theta}{2} + 17.69\cos^4\frac{\theta}{2} - 11.2\cos^6\frac{\theta}{2} + (3\cos^2\varphi - 1)\sin^4\frac{\theta}{2}\left(3.86 - 6.72\cos^2\frac{\theta}{2}\right)$$

$$N(0)T_{t_I} = \sin^2\frac{\theta}{2}\cos\varphi\left[\left(-3.3 + 2.28\cos^2\frac{\theta}{2} - 5.82\cos^4\frac{\theta}{2}\right) - 0.74\sin^4\frac{\theta}{2}(5\cos^2\varphi - 3)\right]$$

$$N(0)T_{t_{II}} = -\left[4.78 + \theta^2\left(0.54\cos^2\frac{\varphi}{2} - 3.05\sin^2\frac{\varphi}{2}\right)\right], \quad N(0)T_{t_{III}} = \left[4.78 + \theta^2\left(0.54\sin^2\frac{\varphi}{2} - 3.05\cos^2\frac{\varphi}{2}\right)\right]$$

(8)

It is noted that in obtaining the above equations we truncate the sum in Eq. (5) for k = 3 at melting pressure.

Now we proceed to calculate quasiparticle relaxation rate. In general we may define the relaxation rate as [3]

$$\tau_{p_1}^{-1} = \sum_{2,3,4} W(\vec{p}_1,...,\vec{p}_4)\delta(\vec{p}_1 + \vec{p}_2 - \vec{p}_3 - \vec{p}_4)\delta(\varepsilon_1 + \varepsilon_2 - \varepsilon_3 - \varepsilon_4)\delta_{\sigma_1+\sigma_2,\sigma_3+\sigma_4}\nu_2^0(1-\nu_3^0)(1-\nu_4^0) \quad (9)$$

or

$$\tau_{p_1}^{-1} = \frac{(m^*)^3}{(2\pi)^6}\int \frac{W_{22}\sin\theta}{\cos\frac{\theta}{2}}d\theta d\varphi d\varphi_2 d\varepsilon_2 d\varepsilon_3 d\varepsilon_4 \delta(\varepsilon_1+\varepsilon_2-\varepsilon_3-\varepsilon_4)\nu_2^0(1-\nu_3^0)(1-\nu_4^0) \quad (10)$$

where $\nu^0$ is the equilibrium Fermi distribution function. Here we can think about two kinds of the quasiparticle relaxation rates, one for Bogoliubov quasiparticle, $\tau_{p_1\uparrow}^{-1}$, and another for normal quasiparticle, $\tau_{p_1\downarrow}^{-1}$. For $\tau_{p_1\uparrow}^{-1}$ and $\tau_{p_1\downarrow}^{-1}$, the function $W_{22}$ in Eq. (10) stands for $(1/4)W_{22}(\uparrow\uparrow) + (1/2)W_{22}(\uparrow\downarrow)$ and $(1/4)W_{22}(\downarrow\downarrow) + (1/2)W_{22}(\downarrow\uparrow)$ respectively. Of course, it is noted that $W_{22}(\downarrow\downarrow)$ relates to quasiparticles interaction of normal state whereas $W_{22}(\uparrow\uparrow)$ is entirely associated with Bogoliubov quasiparticles. The integral of quasiparticle relaxation rate consisting of $W_{22}(\uparrow\uparrow)$, considering its dependence on $\theta$, gives temperature dependence as $T^2$. The value of integral of $W_{22}(\uparrow\downarrow)$ at low temperatures is constant. Hence allow $W_{22}(\uparrow\uparrow)$ to be overlooked in comparison with $W_{22}(\uparrow\downarrow)$ due to the smallness of $W_{22}(\uparrow\uparrow)$. However, this is not the case with $W_{22}(\downarrow\downarrow)$ since the integral consisting of it is constant and as a result integrals consisting of $W_{22}(\downarrow\downarrow)$ give different numerical values from those of $W_{22}(\downarrow\uparrow)$. Accordingly, considering the mentioned information about $W_{22}$, quasiparticle relaxation rate for the superfluid component will be less than that of the normal component. The above formulae may be written as, for example for $\tau_{p_1\downarrow}$,

$$\tau_{p_1\downarrow}^{-1} = \frac{(m^*)^3}{4(2\pi)^5}\int \frac{\sin\theta}{\cos\frac{\theta}{2}}d\theta d\varphi d\varphi_2 \left(|T_{t_I}|^2 d\varepsilon_2 d\varepsilon_3 d\varepsilon_4 \delta(\varepsilon_1+\varepsilon_2-\varepsilon_3-\varepsilon_4)\nu_{2\downarrow}^0(1-\nu_{3\downarrow}^0)(1-\nu_{4\downarrow}^0)\right.$$

$$+ (1/2)|T_{s_I} + T_{t_I}|^2 dE_2 dE_3 d\varepsilon_4 \nu_{2\uparrow}^0(1-\nu_{3\uparrow}^0)(1-\nu_{4\downarrow}^0)\delta(\varepsilon_1+E_2-E_3-\varepsilon_4)$$

$$+(1/2)|-T_{s_l}+T_{t_l}|^2\, dE_2 d\varepsilon_3 dE_4 v_{2\uparrow}^0(1-v_{3\downarrow}^0)(1-v_{4\uparrow}^0)\delta(\varepsilon_1+E_2-\varepsilon_3-E_4)) \qquad (11)$$

where $v^0$ is the equilibrium Fermi distribution function. In order to integrate over energies consist of distribution functions in Eq. (11), we replaced $\delta(\varepsilon)$ by $(1/2\pi)\int_{-\infty}^{\infty}\exp(-i\varepsilon z)dz$. Then we deal with such an integral [4] $(1/2\pi)\int_{-\infty}^{\infty}\exp(-i\varepsilon z)(\pi iT/\sinh(\pi zT))^3 dz$ that is equal to $(1/2)(\pi^2 T^2+\varepsilon^2)/(1+e^{-\varepsilon/T})$.

Integrating over angles in Eq. (11) is trivial, and then we have

$$\tau_{p_1\downarrow}^{-1} \simeq \frac{m^{*3}}{N(0)^2} 0.16(\pi^2 T^2+\varepsilon_1^2)/(1+e^{-\varepsilon_1/T})\ ,\quad \tau_{p_1\uparrow}^{-1} \simeq \frac{m^{*3}}{N(0)^2} 0.15(\pi^2 T^2+E_1^2)/(1+e^{-E_1/T}) \qquad (12)$$

At low temperatures, we have $E_1 \simeq \varepsilon_1 \simeq T$ [3], then both $\tau_{p_1\downarrow}^{-1}$ and $\tau_{p_1\uparrow}^{-1}$ are proportional to $T^2$.

Now we are in a position to consider collision integral of Boltzmman equation in the form of $-i\delta v'_{p,\sigma}\tau_{p,\sigma}^{-1}+I_{p,\sigma}^{in}$ [4]. $\delta v'_{p,\sigma}$ is the deviation of $\delta v_{p,\sigma}$ from local equilibrium, $\delta v_{p,\sigma}^{loc}$, and we have $\delta v'_{p,\sigma} = \delta v_{p,\sigma} - \delta v_{p,\sigma}^{loc} = \delta v_{p,\sigma} - v'_{p,\sigma}\delta E_p$. $v'_{p,\sigma}$ is $\partial v_{p,\sigma}^0/\partial E_p$ where $v_{p,\sigma}^0$ is the equilibrium Fermi distribution function and linearized distribution function $\delta v_{p,\sigma}$ is $v_{p,\sigma}-v_{p,\sigma}^0$. The first and second terms of collision integral are outscattering and inscattering collision integral respectively. The main features of this collision integral, such as its anisotropy and its energy and temperature dependences, are contained in the outscattering term. This term describes relaxation of the distribution function $\delta v_{p,\sigma}$ with a rate $\tau_{p,\sigma}^{-1}$. Boltzmann equation in momentum space subject to an external disturbance of wave vector $q$ for time-independent transport phenomena is [5,6] (for simplicity we omit subscript 1 in $p_1$)

$$-\vec{q}\vec{\nabla}_{\vec{p}} E_{\vec{p}} \delta v'_{p,\sigma} = -i\delta v'_{p,\sigma}\tau_{p,\sigma}^{-1}+I_{p,\sigma}^{in} \qquad (13)$$

where the group velocity of Bogoliubov quasiparticle, $\vec{\nabla}_{\vec{p}} E_{\vec{p}}$, is $(\varepsilon_p/E_p)(p/m^*)$ and $m^*$ is effective mass of $^3He$. After substituting $\delta v'_{p,\sigma}$ in left hand side of Eq. (13) by $\delta v_{p,\sigma}-v'_{p,\sigma}\delta E_p$, we have

$$-\vec{q}\vec{\nabla}_{\vec{p}} E_{\vec{p}} \left(\delta v_{p,\sigma}-v'_{p,\sigma}\delta E_p\right) = -i\delta v'_{p,\sigma}\tau_{p,\sigma}^{-1}+I_{p,\sigma}^{in} \qquad (14)$$

which is valid for both spin-symmetric and spin-antisymmetric distribution functions $\delta v_{p,\sigma}$. In the case of a stationary transport situation it is sufficient to approximate $\delta v_{p,\sigma}$ on the left hand side of Eq. (14) by the local equilibrium distribution in the rest frame of the moving fluid [7]

$$\delta v_{p,\sigma} = v'_{p,\sigma}\left(\delta E_p - \delta E_p^{ext}\right)+O\left(\nabla\cdot V^n\right) \qquad (15)$$

where $\delta E_p^{ext} = p\cdot V^n$ denotes the energy change of a quasiparticle in the normal-fluid velocity field $V^n$. Then we have the following inhomogeneous integral equation for $\delta v'_{p,\sigma}$

$$\frac{\varepsilon_p}{E_p}\frac{p_M q_M}{m^*} p_o V_o^n v'_{p,\sigma} = -i\delta v'_{p,\sigma}\tau_{p,\sigma}^{-1}+I_{p,\sigma}^{in} \qquad (16)$$

The inscattering collision integral, $I_{p,\sigma}^{in}$, at relaxation-time approximation [3,6], is written as

$$I_{p,\sigma}^{in}=i\lambda_{2\sigma}^{+}\tau_{p,\sigma}^{-1}\sum_{m=-2}^{2}\left(v'_{p,\sigma} Y_{2m}(\hat{p})\sum_{p'}\delta v'_{p',\sigma}\tau_{p',\sigma}^{-1} Y_{2m}^{*}(\hat{p}')\Big/\sum_{p'}v'_{p',\sigma}\tau_{p',\sigma}^{-1}|Y_{2m}(\hat{p}')|^2\right) \qquad (17)$$

where $Y_{lm}(\hat{p})$ is spherical harmonic function. Also $\lambda_{2\sigma}^{+}$ is defined as

$$\lambda_{2\sigma}^{+}=1-3\langle W_{22}\sin^4(\theta/2)\sin^2\varphi\rangle/\langle W_{22}\rangle \qquad (18)$$

with $\langle A\rangle \equiv \int(d\Omega/4\pi)(A(\theta,\varphi)/\cos(\theta/2))$. $\sigma$ stands for $\uparrow$ and $\downarrow$. By introducing function $\varphi_{p,\sigma}$ through

$$\delta v'_{p,\sigma} = (1/m^*)(\varepsilon_p/E_p) i v'_{p,\sigma} q_M p_M p_o V_o{}^n \varphi_{p,\sigma} \tag{19}$$

and then after substituting $\delta v'_{p,\sigma}$, given by Eq. (19), in Eq. (16) and using Eq. (17), then we have

$$\tau_{p,\sigma} = \varphi_{p,\sigma} - \lambda^+_{2\sigma} \sum_{m=-2}^{2} \left( \frac{Y_{2m}(\hat{p}) \sum_{p'} q_M p'_M V_o{}^n p'_o v'_{p',\sigma} \tau^{-1}_{p',\sigma} Y^*_{2m}(\hat{p}') \varphi_{p',\sigma}}{q_M p_M V_o{}^n p_o \sum_{p'} v'_{p',\sigma} \tau^{-1}_{p',\sigma} |Y_{2m}(\hat{p}')|^2} \right) \tag{20}$$

Furthermore, we can obtain $\varphi_{p,\sigma}$ from Eq. (20) as follows.

$$\varphi_{p,\sigma} = \tau_{p,\sigma} + \frac{\lambda^+_{2\sigma}}{1 - \lambda^+_{2\sigma}} \sum_{m=-2}^{2} \frac{Y_{2m}(\hat{p}) \sum_{p'} v'_{p',\sigma} Y^*_{2m}(\hat{p}')}{\sum_{p'} v'_{p',\sigma} \tau^{-1}_{p',\sigma} |Y_{2m}(\hat{p}')|^2} \tag{21}$$

By substituting $\delta v'_{p,\sigma}$, given by Eq. (19), in the following equation of the momentum flux tensor in momentum space

$$\Pi_{ij} = (1/m^*) \sum_{p,\sigma} (\varepsilon_p/E_p)(p_i p_j - (1/3)\delta_{ij} p_F p_F) \delta v'_{p,\sigma} \tag{22}$$

and finally comparing Eq. (22) with $\Pi_{ij} = -i\eta_{ijlk} q_l V_k{}^n$, we can drive shear viscosity tensor of superfluid $^3He - A_1$ as follows

$$\eta_{ijlk} = -(p_F^5/m^*) \sum_\sigma \langle\langle \hat{p}_i \hat{p}_j \hat{p}_l \hat{p}_k v'_{p,\sigma} \varphi_{p,\sigma} \rangle\rangle \tag{23}$$

where $p_F$ is Fermi momentum. For writing Eq. (23), we have used the fact that at low temperatures all momenta lie on the Fermi surface. After substituting $\varphi_{p,\sigma}$, given by Eq. (21), in Eq. (23), we have

$$\eta_{ijlk} = -\left(\frac{p_F^5}{m^*}\right) \sum_\sigma \left[ \langle\langle \hat{p}_i \hat{p}_j \hat{p}_l \hat{p}_k v'_{p,\sigma}/\tau^{-1}_{p,\sigma} \rangle\rangle + \frac{\lambda^+_{2\sigma}}{1-\lambda^+_{2\sigma}} \sum_{m=-2}^{2} \left( \frac{\langle\langle \hat{p}_i \hat{p}_j v'_{p,\sigma} Y_{2m}(\hat{p}) \rangle\rangle \langle\langle \hat{p}_l \hat{p}_k v'_{p,\sigma} Y^*_{2m}(\hat{p}) \rangle\rangle}{\langle\langle v'_{p,\sigma} \tau^{-1}_{p,\sigma} |Y_{2m}(\hat{p})|^2 \rangle\rangle} \right) \right] \tag{24}$$

where $\langle\langle A \rangle\rangle \equiv (2/(2\pi)^3) \int d\Omega_p \int_{-\infty}^{\infty} d\varepsilon_p A(\theta,\varphi)$ with $d\Omega_p = d\Omega$, considering the direction of initial quasiparticle momentum oriented toward z-axis. In Eq. (24), just $v'_{p,\sigma}$ and $\tau^{-1}_{p,\sigma}$ (see Eq. (12)) are functions of $\varepsilon_p$ and integrating over $\varepsilon_p$ is trivial. After considering $\hat{p}_x = \sin\theta\cos\varphi$, $\hat{p}_y = \sin\theta\sin\varphi$, $\hat{p}_z = \cos\theta$ and the obtained values of $\lambda^+_{2\sigma}$ ($\lambda^+_{2\downarrow} \simeq \lambda^+_{2\uparrow} \simeq 0.75$) and also performing integration over $d\Omega_p$, Finally we obtain

$$\eta_{xy} = \eta_{xz} = \frac{1}{3}\eta_{zz} \simeq \frac{N(0)^2 p_F^5}{m^{*4}} 0.02 \frac{1}{T^2} \tag{25}$$

where we consider that $\eta_\uparrow$ is negligible with respect to $\eta_\downarrow$ in total shear viscosity, $\eta (\eta = \eta_\downarrow + \eta_\uparrow)$. It is noted that superfluid components do not play a role in obtaining shear viscosity tensor, because the temperature dependence of these components are proportional to $T^4$, $T^2$ and $T^0$ for the components $\eta_{xy}$, $\eta_{xz}$ and $\eta_{zz}$, respectively. For example to obtain temperature dependence of $\eta_{xy\uparrow}$, by considering smallness of $\theta$, we have used the relations

$$\int_{-\infty}^{\infty} v'_{p,\sigma}/\tau^{-1}_{p,\sigma} d\varepsilon_p = (-3.33/T^2)(N(0)^2/m^{*3})$$

and

$$\int \hat{p}_x \hat{p}_y \hat{p}_x \hat{p}_y d\Omega = \int_0^{\pi T/1.77T_c} \theta^5 d\theta \int_0^{2\pi} \sin^2\varphi \cos^2\varphi d\varphi = (0.13)(\pi/1.77)^6 (T/T_c)^6 \ .$$

By multiplication of these two integrals (and considering the nodes of energy gap), it is seen that the first term in the bracket of $\eta_{xy\uparrow}$ in Eq. (24) is proportional to $T^4$. Also we have used the following relations for the calculation of the second term in the bracket of $\eta_{xy\uparrow}$

$$c \equiv \int_{-\infty}^{\infty} v'_{p,\sigma} \tau_{p,\sigma}^{-1} d\varepsilon_p = -0.99 \left(m^{*3}/N(0)^2\right) T^2 \ , \ b \equiv \int_{-\infty}^{\infty} v'_{p,\sigma} d\varepsilon_p = -1 \ , \ a \equiv \int \hat{p}_x \hat{p}_y Y_{22}(\hat{p}) d\Omega_p \ ,$$

$$a^* \equiv \int \hat{p}_x \hat{p}_y Y_{22}^*(\hat{p}) d\Omega_p = -i(0.10)(\pi/1.77)^6 (T/T_C)^6 \ \text{and} \ d \equiv \int Y_{22}(\hat{p}) Y_{22}^*(\hat{p}) d\Omega_p = 0.94(\pi/1.77)^6 (T/T_C)^6 .$$

It is seen that $aa^*b^2/cd\pi^3$, which is nonvanishing term $m=2$ of the second term in the bracket of $\eta_{xy\uparrow}$, is proportional to $T^4$. In addition, nonvanishing term $m=-2$ is equal to the term $m=2$. Therefore $\eta_{xy\uparrow}$ is proportional to $T^4$.

Now we proceed to express Eq. (25) in terms of the shear viscosity at $T_c$, $\eta(T_c)$. $\eta(T_c)$ is $(1/5)\rho(m^*/m)v_F^2\tau_0$ with $\tau_0$ is a characteristic relaxation time and is given by $8\pi^4/m^{*3}T_c^2\langle W_N \rangle$ and density of the liquid, $\rho$, is $mp_F^3/3\pi^2$ and $\langle W_N \rangle = (2\pi/8)\int d\Omega \left(3|T_{t_1}|^2 + |T_{s_1}|^2\right)/4\pi \cos\frac{\theta}{2} \simeq 130.56/N(0)^2$. Then we have the following formulae that is the main result of this paper

$$\eta_{xy} = \eta_{xz} = (1/3)\eta_{zz} \simeq 0.50(T/T_c)^{-2}\eta(T_c) \tag{26}$$

In conclusion, for superfluid $^3He$-$A_1$, we calculate the transition probabilities for the cases where both the normal and Bogoliubov quasiparticles are presented in decay, coalescence and binary processes, beyond s-p approximation by using Pfitzner procedure. It is noted that this is the same procedure applied for the results near $T_c$ [8]. It is interesting to mention that interaction between superfluid and normal fluid in the calculation of the transition probabilities is distinct, whereas the contribution from the interaction between Bogoliubov quasiparticles at low temperatures is negligible. As a result, the integral consisting of $W_{22}(\uparrow\uparrow)$ is negligible with respect to that of $W_{22}(\uparrow\downarrow)$. Of course, this can be considered as one of the results of the paper. Also it is mentioned that just binary processes are dominated in the calculation of the transition probabilities.

Then we calculate quasiparticle relaxation rates at low temperatures and in high magnetic field at melting pressure. Our results show that quasiparticle relaxation rates are proportional to $T^2$ for both normal and superfluid components, but there is a slight difference between the two prefactors due to difference of $W_{22}$ for normal and Bogoliubov quasiparticles. This can be considered as another result of the paper. Our results of quasiparticle relaxation rates can be used for calculation of other transport coefficients of superfluid $^3He$-$A_1$.

In this paper we use Boltzmann equation in momentum space where collision integral in the equation contains inscattering and outscattering terms. By using this method and the results of quasiparticle relaxation rates, we calculate the components of shear viscosity tensor of superfluid $^3He$-$A_1$. We obtain that all of the components $\eta_{xy}$, $\eta_{xz}$ and $\eta_{zz}$ are proportional to $T^{-2}$. In addition, we obtain $\eta_{xz} \simeq 0.50(T/T_c)^{-2}\eta(T_c)$ (see Eq. (26)). This can be considered as another result of the paper. Our result is in fairly good agreement with the experiments of Roobol et al. It should be noted that Roobol et al's [9] results on the shear viscosity of superfluid $^3He$-$A_1$ in a static magnetic field up to 15 T at low temperatures indicate $\eta = 0.48\eta(T_c)(T/T_c)^{-2}$, which $\eta_{xz}$ was mainly measured with a few percent contribution of $\eta_{zz}$. It is seen that our prefactor of $\eta$ has a little larger value than Roobol et al's prefactor. It is also mentioned that temperature dependences of the components $\eta_{xy\uparrow}$, $\eta_{xz\uparrow}$ and $\eta_{zz\uparrow}$ are $T^4$, $T^2$ and $T^0$, respectively. These results have also been calculated by us [10] at low temperatures by using Sykes et al's [11] method, which are the same as those mentioned above. Roobol et al by using the method of temperature asymptotes obtained these temperature dependences, which are one order less than our results. However, in any case, these components are negligible in comparison to the normal component, $\eta_\downarrow$. To calculate $\eta_\downarrow$, by some method one should obtain singlet and triplet QSA. It is noted that s- approximation, s-p approximation, forward scattering approximation and so on [6,12] have been used to express the singlet and triplet QSA, although we here (and in [8,10]) have used Pfitzner procedure to calculate them. Of course, for a system as dense as $^3He$ there is no physical reason to expect the contribution from higher partial waves in QSA to be negligible. Whereas in Pfitzner procedure, necessary conditions are established to explicitly construct exchange-symmetric scattering amplitudes by adding higher angular momentum components. In this procedure, a general polynomial expansion of the QSA is constructed. This expansion fully contains the s-p approximation as a special case (truncated expansion at $k=1$ in Eq. (5)). Therefore Pfitzner procedure improves other mentioned approximations. Roobol et al [9] proposed $\eta_\downarrow = 0.473\eta(T_c)(T/T_c)^{-2}$ based on Fermi liquid model. Presumably they used s-wave approximation in their calculations, whereas we use Pfitzner procedure. The minor discrepancy between our prefactor of $\eta$ and Roobol et al's, 0.473, arises mainly from improving the calculation of singlet and triplet QSA via Pfitzner procedure by us. Anyway, by the method applied in this paper along with using Pfitzner procedure, we theoretically obtain 0.5 for prefactor of $\eta$. Furthermore, our result [10] of the shear viscosity obtained by using Sykes et al's [11] method along with using Pfitzner procedure is nearly the same as the results of this paper.